\documentclass[conference]{IEEEtran}
\IEEEoverridecommandlockouts
\usepackage{cite}
\usepackage{amsmath,amssymb,amsfonts}
\usepackage{algorithmic}
\usepackage{graphicx}
\usepackage{textcomp}
\usepackage{xcolor}
\def\BibTeX{{\rm B\kern-.05em{\sc i\kern-.025em b}\kern-.08em
    T\kern-.1667em\lower.7ex\hbox{E}\kern-.125emX}}
\begin{document}

\title{Open Set Wireless Standard Classification Using Convolutional Neural Networks\\
\thanks{The support of the US Army Research Laboratory is gratefully acknowledged.}
}
\author{
\IEEEauthorblockN{Samuel R. Shebert}
\IEEEauthorblockA{\textit{Bradley Department of ECE} \\
\textit{Virginia Tech}\\
Blacksburg, USA \\
sshebert@vt.edu}
\and
\IEEEauthorblockN{Anthony F. Martone}
\IEEEauthorblockA{\textit{US Army Research Laboratory} \\
Adelphi, USA \\
anthony.f.martone.civ@mail.mil}
\and
\IEEEauthorblockN{R. Michael Buehrer}
\IEEEauthorblockA{\textit{Bradley Department of ECE} \\
\textit{Virginia Tech}\\
Blacksburg, USA \\
rbuehrer@vt.edu}
}

\maketitle

\begin{abstract}
In congested electromagnetic environments, cognitive radios require knowledge about other emitters in order to optimize their dynamic spectrum access strategy. 
Deep learning classification algorithms have been used to recognize the wireless signal standards of emitters with high accuracy, but are limited to classifying signal classes that appear in their training set. 
This diminishes the performance of deep learning classifiers deployed in the field because they cannot accurately identify signals from classes outside of the training set.
In this paper, a convolution neural network based open set classifier is proposed with the ability to detect if signals are not from known classes by thresholding the output sigmoid activation. 
The open set classifier was trained on 4G LTE, 5G NR, IEEE 802.11ax, Bluetooth Low Energy 5.0, and Narrowband Internet-of-Things signals impaired with Rayleigh or Rician fading, AWGN, frequency offsets, and in-phase/quadrature imbalances. 
Then, the classifier was tested on OFDM, SC-FDMA, SC, AM, and FM signals, which did not appear in the training set classes.
The closed set classifier achieves an average accuracy of 94.5\% for known signals with SNR's greater than 0 dB, but by design, has a 0\% accuracy detecting signals from unknown classes.
On the other hand, the open set classifier retains an 86\% accuracy for known signal classes, but can detect 95.5\% of signals from unknown classes with SNR's greater than 0 dB.
\end{abstract}


\section{Introduction}
Congested electromagnetic environments (EME) are a challenging scenario for communication and radar devices. Cognitive radios and radars have been implemented to improve performance is these complex environments. However, improvements are limited without knowledge about other signals in the environment. Cognitive radios need the ability to identify unknown signals to be able to choose an appropriate dynamic spectrum access (DSA) policy. 

In previous works, feature-based classifiers have been presented to determine the wireless standard of unknown signals \cite{Li2019}. These classifiers use features such as bandwidth, center frequency, synchronization signals, or cyclostationary properties present in the signals. In \cite{FuzzyLogic}, features like bandwidth and frequency hopping characteristics were used with a fuzzy logic classifier to recognize signal standards. In \cite{alfi2019}, detection of generalized frequency division multiplexing (GFDM) signals was accomplished using the cyclostationary properties of the signals. The authors in \cite{8108218} use a combination of bandwidth, synchronization signals specific to certain standards, and modulation index to identify signals in the ISM band.

These feature-based approaches rely on feature design to distinguish signals. This is when an expert determines features to discriminate between the classes \cite{Li2019}. In practice, it can be difficult to identify clear features that differentiate classes. This has led to the use of feature-learning approaches, which can learn relevant features using datasets of signals without any expert design \cite{Li2019}. The feature-learning capabilities, combined with high classification accuracy, have made deep learning approaches a popular option for wireless standards identification (WSI).


Table \ref{tab:classifier_review} gives an overview of previous works in the area of deep learning-based WSI. Deep neural networks (DNN) and convolutional neural networks (CNN) are commonly used deep learning architectures.

\begin{table}[h]
    \caption{Previous Deep Learning-Based WSI Works}
    \label{tab:classifier_review}
    \begin{tabular}{ p{\dimexpr 0.15\linewidth-2\tabcolsep} p{\dimexpr 0.15\linewidth-2\tabcolsep} p{\dimexpr 0.35\linewidth-2\tabcolsep} p{\dimexpr 0.35\linewidth-2\tabcolsep} }
        \hline
        Author & Model & Standards & Impairments\\
        \hline
        \cite{Alshathri2019} & DNN & GSM, UMTS, LTE & AWGN\\
        \hline
        \cite{Alhazmi2020} & CNN & UMTS, LTE, 5G & AWGN, Rayleigh Fading\\
        \hline
        \cite{Xia2019} & CNN & GSM, UMTS, LTE & AWGN, Rayleigh Fading\\
        \hline
        \cite{hiremath2018} & CNN & GSM, Bluetooth, WiFi & AWGN\\
        \hline
        \cite{Ahmed2017} & CNN & WLAN, LTE, Radar & AWGN\\
        \hline
        \cite{Naim2017} & CNN & WiFi, Bluetooth, Zigbee & AWGN\\
        \hline
        \cite{Merima2017} & CNN & WiFi, Bluetooth, Zigbee & AWGN, Flat Fading\\
        \hline
        \cite{Behura2020} & CNN & UMTS, LTE, WiFi, Bluetooth, +24 more & AWGN\\
        \hline
    \end{tabular}
\end{table}

\begin{table*}[t]
\begin{center}
    \caption{Overview of Signal Standard Parameters}
    \label{tab:signal_standards}
    \begin{tabular}{ c  c  c  c  c  c  c  c }
        \hline
        & 4G Downlink & 5G Downlink & 4G Uplink & 5G Uplink & WiFi 6 & BLE 5.0 & NB-IoT\\
        \hline
        Waveform & OFDMA & OFDMA & SC-FDMA & OFDMA, SC-FDMA & OFDMA & GFSK & OFDMA\\
        \hline
        Bandwidth (MHz) & 1.4 -- 20 & 5 -- 100 & 1.4 -- 20 & 5 -- 100 & 20 -- 80 & 2 & 0.18 \\
        \hline
        Subcarrier Spacing & 15 kHz & 15, 30 kHz & 15 kHz & 15, 30 kHz & 78.125 kHz & N/A & 15 kHz\\
        \hline
        Modulation & QPSK--64QAM & QPSK--256QAM & QPSK--64QAM & BPSK--64QAM & BPSK--1024QAM & GFSK & QPSK\\        
        \hline
        Duplex Mode & FDD & FDD & FDD & FDD & TDD & TDD & FDD\\
        \hline
        Devices & 1 & 1 & 1-8 & 1-14 & 1-9 & 1-14 & 1\\
        \hline
    \end{tabular}
\end{center}
\end{table*}

However, a major drawback to DNN and CNN models is their inability to appropriately classify signals that the model was not pretrained to classify. In practice, the large amount of existing signal standards, combined with the perpetual development of new standards makes it impractical to include all signal classes in the training data. Therefore, it is reasonable to assume that deployed deep learning-based WSI systems would need to identify signals they were not pretrained to classify. 


In this paper, an open set CNN-based wireless standard classifier is proposed. This improves upon previous WSI classifiers by adding an unknown class detection algorithm to identify signals from unknown classes. This prevents the deep learning classifier from missclassifying these signals. The unknown class detection algorithm developed by \cite{Shu2017} is adapted and applied to our CNN-based signal classifier. 

The CNN classifier is trained using the 4G LTE downlink and uplink, 5G NR downlink and uplink, IEEE 802.11ax (WiFi 6), Bluetooth Low Energy (BLE) 5.0, and Narrowband Internet-of-Things (NB-IoT) standards. These signals are simulated with Rayleigh or Rician fading, AWGN, frequency offsets, and in-phase/quadrature (I/Q) imbalances to make the signals more realistic. The open set capabilities are tested on a set of unknown signal classes made up of orthogonal frequency-division multiplexing (OFDM), single carrier frequency-division multiple access (SC-FDMA), single carrier (SC), amplitude modulation (AM), and frequency modulation (FM) signals.

\section{Open Set Classification}
In general, there are four categories of classes in classification problems: known known, known unknown, unknown known, and unknown unknown \cite{Geng2020}. Two of these categories are relevant for open set classification: known known and unknown unknown classes. Known known classes are the labeled classes used to train the model, and will be referred to as "known classes". Unknown unknown classes are classes that the model has no information about during training, and will be referred to as "unknown classes" \cite{Geng2020}.

Open set classification approaches can be separated into two main areas, generative and discriminative. In generative approaches, the classifier is trained to classify unknown classes as an additional class \cite{Geng2020}. This means that an $N$ class model would use $N+1$ classes with the final class representing all unknown classes. With this method, the training data for the unknown class must be carefully chosen, and generative machine learning models can assist in this process. On the other hand, discriminative approaches use an algorithm external to the classifier to identify samples from unknown classes. Both approaches are viable for open set classification, but discriminative algorithms have been more frequently chosen by researchers \cite{Geng2020}. 

Bendale and Boult developed a discriminative algorithm in the form of a new layer for deep learning models, called OpenMax \cite{Bendale2016}. This layer acts as unknown class detector for open-set images or adversarially generated images. The authors of \cite{Shu2017} propose an alternative to OpenMax by using thresholded Sigmoid values to reject unknown signal classes and found that their method outpreformed OpenMax for text classification. In this paper, the unknown class detector proposed by the authors in \cite{Shu2017} is applied to wireless standards classification.


\section{Dataset Generation}
All signal datasets used for training and testing were generated in MATLAB. The known classes used MATLAB 2021a and the unknown classes used MATLAB 2018b. For each wireless standard, 1440 10 ms signals were generated, corresponding with the 10 ms frame duration of the 4G LTE and 5G NR standards. Signals were generated using random transport data and with different sets of parameters, so that the dataset included a wide range of options found in the standards. An overview of the parameters can be found in Table \ref{tab:signal_standards}. 

Since many signal bandwidths were included in the dataset, the Nyquist sampling rate for each bandwidth differs. However, CNN-based models require a fixed number of samples per classification. Therefore, all signals were upsampled to a 125 MHz sampling rate to ensure the minimum sampling requirements of all signals were met.

A brief outline of the known and unknown signal classes will be discussed next. The known class datasets were labeled data used to train the classifier. The unknown class datasets were not exposed to the classifier during training, and were used to test the unknown class detection capabilities. 7 known classes and 5 unknown classes were used.

\subsection{Known Classes}
\subsubsection{4G LTE \& 5G NR}
4G LTE downlink and uplink signals were generated to be release 8 compliant using the MATLAB LTE Toolbox. 5G NR downlink and uplink signals were generated to be release 15 compliant using the MATLAB 5G Toolbox. Resource block utilization of the shared channels ranged from 1\% to 100\% and the allocation of resource blocks followed two traffic models. The first traffic model randomly allocated resource blocks throughout the radio frame. The second model replicated bursty traffic by allocating large contiguous blocks of resource blocks sporadically in time.

\subsubsection{IEEE 802.11ax (WiFi 6)}
WiFi 6 signals were generated using the MATLAB WLAN Toolbox. The single-user, extended-range single-user, multi-user, and trigger based packet formats were included in the dataset. Randomized traffic was simulated by using packets with different transport data lengths and inserting periods of no transmission between packets.

\subsubsection{Bluetooth Low Energy (BLE) 5.0}
BLE signals were generated using the MATLAB Communications Toolbox Library for the Bluetooth Protocol. The BLE signals included simulated advertisers, pairing devices, and ongoing connections.

\subsubsection{Narrowband Internet-of-Things (NB-IoT)}
NB-IoT downlink signals were generated to be release 13 compliant using the MATLAB LTE Toolbox. The NB-IoT signals were generated in stand-alone mode, as opposed to the guard-band or in-band modes. 

\subsection{Unknown Classes}
\subsubsection{OFDM, SC-FDMA, \& SC}
OFDM, SC-FDMA, and SC waveforms are used in many of the aforementioned standards in the dataset of known signals. However, the signals used in the outlier set were generic, and lacked the synchronization and control signals, as well as any higher level structures found in the standards.

Signal modulation was varied between QPSK, 16PSK, 64PSK, 16QAM, 64QAM, and 256QAM and signal bandwidth was varied between 25-100 MHz. The subcarrier spacings of the OFDM and SC-FDMA signals was 15, 30, or 60 kHz.

\subsubsection{AM \& FM}
The AM and FM signals were the only analog modulations used in the known and unknown classes. Single and double side-band AM signals were used.

\section{Dataset Processing}
To improve the accuracy of the CNN classifier, data augmentation and signal preprocessing were performed. Data augmentation was used to make the training signals more realistic and promote the CNN classifier's generalizability. Signal preprocessing was applied to all signals, training and testing, to prepare the signals to be input into the CNN.

\begin{figure}
    \centering
    \includegraphics[scale=0.47]{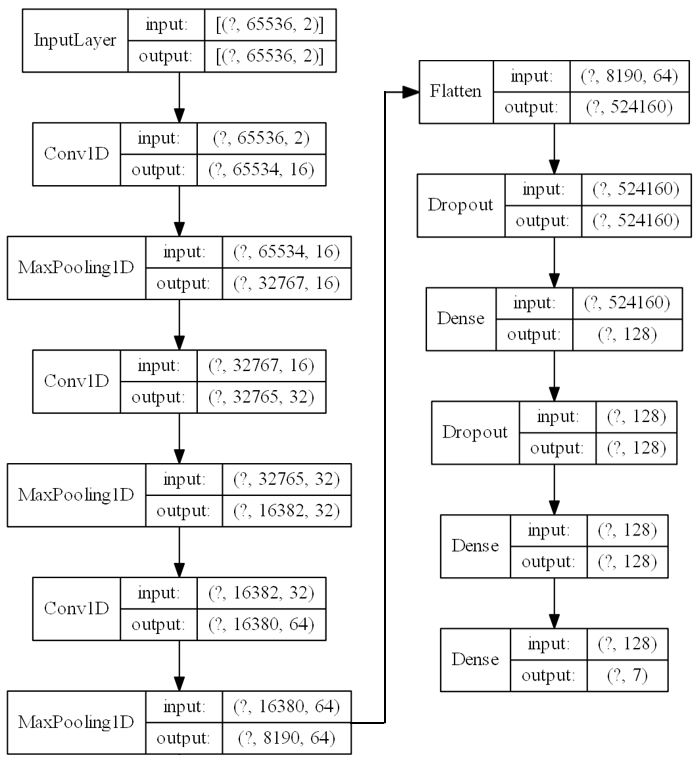}
    \caption{CNN architecture, question marks are a placeholder for batch size}
    \label{fig:modelArchitecture}
\end{figure}

\subsection{Data Augmentation}
The first data augmentation step was to randomly sample portions of each signal with replacement, in a process similar to bootstrap sampling. The purpose of this step was to generate shorter signals that matched the input size of the CNN and ensure that the training signals were not time-synchronized with any structure of the standards. A duration of 1 ms per random sample was determined experimentally to be long enough for the model to learn relevant features to detect unknown classes, while being small enough to remain computationally tractable. The 1440 10 ms signals for each class were randomly sampled 10 times, resulting in 14400 subsignals per class. An 80, 10, 10 split was used for the training, validation, and test datasets for the known classes. This meant that 90720 training and validation and 10108 1 ms testing signals were used.

The next data augmentation step added common wireless signal impairments to prepare the CNN model for typical conditions in wireless communication. Multipath fading, I/Q imbalances, frequency offsets, and AWGN impairments were used. First, each training signal was impaired by Rayleigh or Rician multipath fading. Then, the real part of each signal was randomly scaled by -3 to 3 dB to simulate I/Q imbalances. Next, a random frequency offset in the range of -2500 to 2500 Hz was added. Finally, AWGN ranging from -10 to 20 dB SNR was added. The AWGN used in the training set was limited to -10 dB because it was found that training with lower SNR signals degraded accuracy across all SNR levels.

\subsection{Signal Preprocessing}
Three preprocessing steps were used: (1) the Short-Term Fourier Transform (STFT); (2) the magnitude of the STFT; (3) the normalization of the resulting magnitude. These preprocessing steps were experimentally found to increase the classification accuracy of the model. The STFT function separated a 131072 sample signal, corresponding to approximately 1 ms of data at a 125 MHz sampling rate, into two, 65536, sample signals. Then the Fast Fourier Transform (FFT) of both segments was taken. The CNN model was able to achieve higher accuracy using the time-frequency domain STFT, over time-domain I/Q samples or frequency-domain FFTs. 

Next the magnitude was taken to convert the complex-valued signal to real-valued. This had a neglible impact on classifier performance, while halving the memory size of each signal. Finally, the signals were normalized between 0 and 1 to limit the range of values exposed to the classifier.

\section{Model Architecture}
\begin{figure}
    \centering
    \includegraphics[scale=0.49]{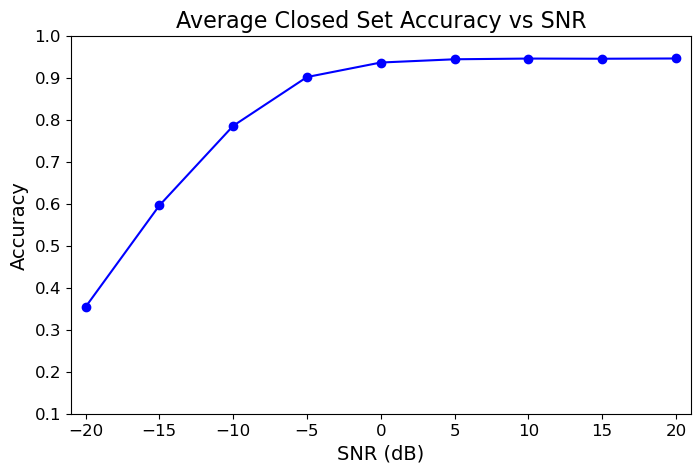}
    \caption{Average closed set accuracy over a range of SNR values. All test signals are from known classes.}
    \label{fig:closedAcc}
\end{figure}

The classifier model was composed of a 1D CNN. The model was developed in Python 3.8.8 using Tensorflow 2.4.1. The unknown detection algorithm built on top of CNN model by thresholding the output.

\subsection{CNN Model}
CNN models are widely used with high dimensional input data, such as images or signals. Convolution and max pooling layers are used at the head of the network to learn relevant features in the data. This is followed by dense layers at the tail of the network, which perform the final classification of the data.

Figure \ref{fig:modelArchitecture} shows the layer types and sizes of the CNN model used. The convolution layers used a kernel size of 3 and a stride of 1. All of the convolution and dense layers, except for the final layer, used ReLu activation. The final dense layer used sigmoid activation, as required by the unknown class detection algorithm. The model was trained for 10 epochs using the categorical cross-entropy loss function, the Adam optimizer, and a batch size of 128.

The models were trained and tested using computational clusters provided by Advanced Research Computing at Virginia Tech. The allocation used for training and testing had 8 CPU cores, 210 GBs of RAM, and 1 Nvidia v100 GPU with 16 GB VRAM.

\begin{figure}
    \centering
    \includegraphics[scale=0.49]{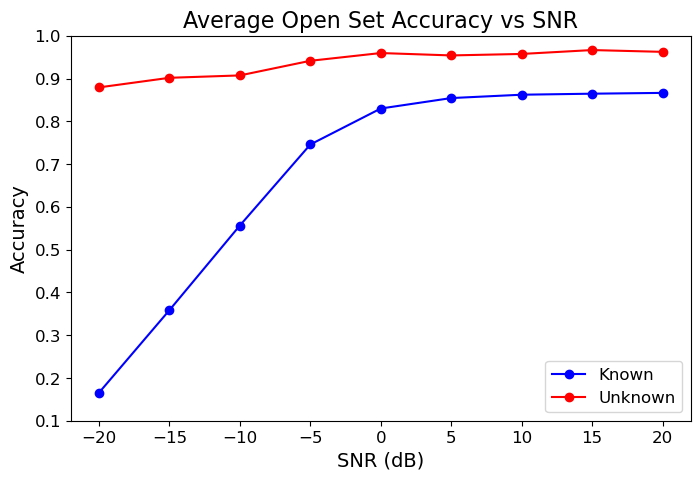}
    \caption{Average open set accuracy over a range of SNR values. Blue signals are known classes and red signals are unknown classes.}
    \label{fig:openAvgAcc}
\end{figure}

\subsection{Unknown Class Detector}
The unknown class detector was used to determine when signals were outside of the training data of the classifier, and therefore could not be accurately classified. This was done by thresholding the output of the sigmoid activation function to identify signals that were too different from the known classes. The effectiveness of a thresholding approach depends on the activation function used at the output of the classifier model. The softmax activation function is commonly used for multi-class classification, but is not suitable for use in open set problems. The softmax function for the $i$th class of an $N$ class classifier is as follows:
\begin{equation}
    \text{softmax} (x_i) = \frac{e^{x_i}}{\sum^N_{j=1}e^{x_j}}
\end{equation}
Where $x$ is a vector of output values from the classifier. The term in the denominator normalizes the output of the softmax function, such that the sum of all of the softmax values will be 1. However, normalization is only reasonable if the classes are collectively exhaustive, such that the probability of all events occuring must sum to 1. This assumption is broken in the open set case, which is premised on the existence of additional classes that are not in the training set. Therefore, the softmax function is not reasonable to use for the open set case.


On the other hand, the sigmoid function does not assume that the classes are collectively exhaustive. The sigmoid function for the $i$th class is as follows:
\begin{equation}
    \text{sigmoid}(x_i) = \frac{1}{1+e^{-x_i}}
\end{equation}
The sigmoid function independently estimates the probability of each class.
If a signal from an unknown class is not similar enough to any of the known classes, the sigmoid function should return lower probabilities. Therefore, a threshold can be used to determine when a signal should be flagged as from an unknown class.

This approach for unknown class detection is a simple and efficient addition to deep learning classifiers, as it requires very little additional processing. The threshold can be tuned empirically using the dataset of unknown classes. The unknown class detection algorithm for an arbitrary signal and classifier with \textit{N} classes is as follows.

\begin{enumerate}
    \item Classify the signal and obtain \textit{N} values, corresponding to the probability of the signal belonging to each class.
    \item For each value, compare to the threshold
    \item If none of the values are above the threshold, the signal is from an unknown class. Otherwise, the signal belongs to the class with the largest probability.
\end{enumerate}

\section{Results}


The average closed set performance of the classifier is shown in Figure \ref{fig:closedAcc}. In the closed set case, the unknown class detection algorithm is not applied and the CNN is only tested with signals from known classes. The classifier plateaus at 94.5\% overall accuracy for SNR values greater than 0 dB. The primary loss of accuracy is due to confusion between the 5G downlink and uplink classes. This could be because in some cases, the 5G downlink and uplink physical layers are nearly identical. 

As the SNR decreases below 0 dB, the accuracy steadily drops to around 35\% accuracy at -20 dB. The accuracy declines  because as the SNR reduces, the signal becomes more corrupted by the AWGN, until the signal is barely recognizable to the classifier at -20 dB.

The closed set classifier is competitive with the performance of other wireless standards classifiers. The authors in \cite{Alhazmi2020} developed a closed set CNN classifiers that achieved 85\% accuracy when SNRs were greater than 0 dB and Rayleigh fading was used. The authors in \cite{Xia2019} achieved 100\% accuracy using AWGN and Rayleigh fading, however 5G signals were not used, which accounted for most of the lost accuracy in our classifier.

\begin{figure}
    \centering
    \includegraphics[scale=0.49]{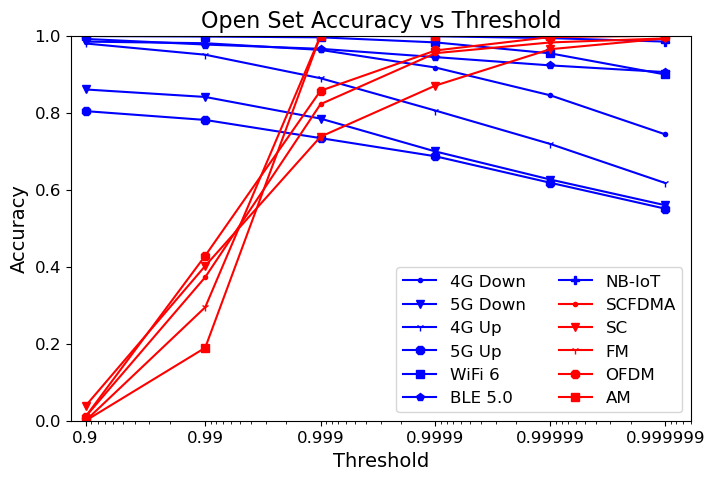}
    \caption{Open set classification of known and unknown signals vs sigmoid threshold. Blue signals are known classes and red signals are unknown classes. SNR of all signals was 10 dB.}
    \label{fig:openThresh}
\end{figure}

The next results show the open set performance of the classifier, which extends previous work done in WSI. In these tests, the classifier was trained with signals from known classes and tested with signals from previously unseen, unknown classes. The unknown signal classes are considered to be correctly classified if they are identified as an unknown class. 

The average open set performance of the classifier is shown in Figure \ref{fig:openAvgAcc}. The open set classifier is able to identify unknown classes with over 95\% accuracy for SNRs greater than 0 dB. The cost of the enhanced capabilities of the open set classifier is a reduced 86\% accuracy for known classes, almost 10\% lower than the closed set case.

For negative SNR values, the accuracy of known classes show a trend that is similar, but exacerbated compared to the closed set case. Conversely, the unknown classes detection accuracy did not deteriorate much with low SNR. This is because very noisy signals are dissimilar to the signals used to train the classifier, so the probabilities at the output of the sigmoid function are lower on average. Therefore, both unknown and known signals should detected as unknown at a higher rate as the SNR drops. Interestingly, the unknown signals decrease slightly in accuracy under high noise. This could be because the high noise is enough to randomly shift some of the unknown signals into the decision regions of the known classes.

Figure \ref{fig:openThresh} shows the classification accuracy over multiple sigmoid thresholds. As the threshold increases, the known signals decrease in accuracy. This is due to the higher thresholds shrinking the decision boundaries of the known classes. On the other hand, increasing the threshold increases the number signals from unknown classes that are properly identified. In general, as the threshold value goes to 0, the accuracy of detecting unknown classes will go to 0\% and the accuracy of the known classes will be equivalent to the closed set case. As the threshold goes to 1, all signals will be flagged as unknown classes. 

The threshold should be tuned to maximize the detection of the unknown classes and minimize the loss of accuracy in the known classes. A threshold of 0.9999 was chosen since it has good unknown class detection without deteriorating the known class accuracies too significantly.

\begin{figure}
    \centering
    \includegraphics[scale=0.49]{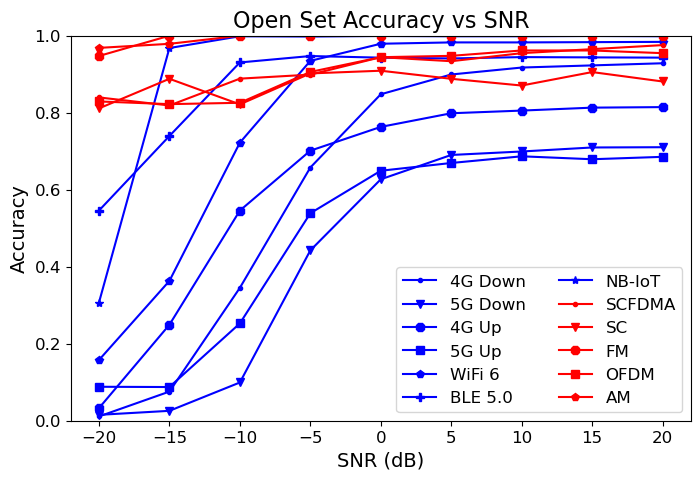}
    \caption{Open set classification of known and unknown signls vs SNR. Sigmoid threshold was set at 0.9999. Blue signals are known classes and red signals are unknown classes.}
    \label{fig:openAcc}
\end{figure}

\begin{figure}
    \centering
    \includegraphics[scale=0.57]{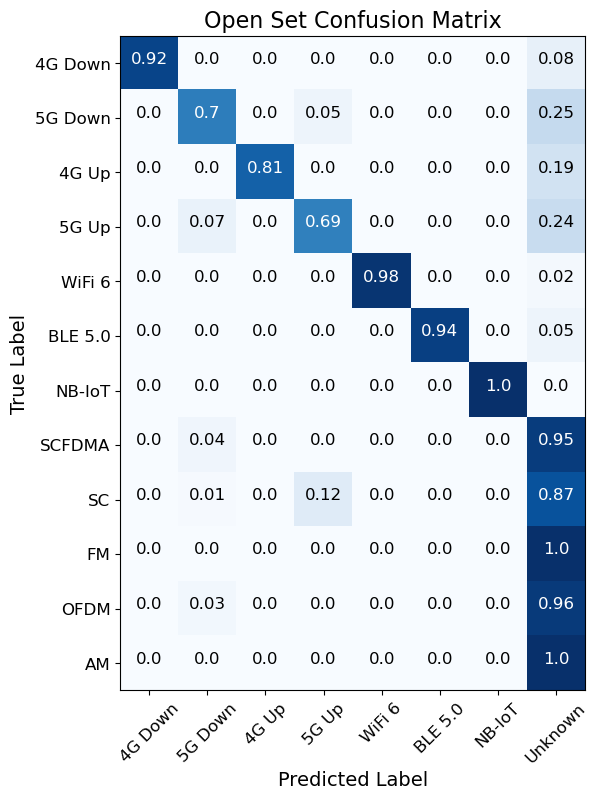}
    \caption{Confusion matrix for open set classifications of known and unknown classes. Sigmoid threshold was set at 0.9999. SNR of all signals was 10 dB}
    \label{fig:openConfusion}
\end{figure}

Figure \ref{fig:openAcc} shows the accuracy for each class of the open set classifier. It can be seen that the accuracy of the known classes that were more difficult to classify in the closed set case, such as the 4G and 5G signals, were more severely affected in the closed set case. A closer examination of this can be seen in Figure \ref{fig:openConfusion}. Out of the known classes, 4G and 5G signals are most often identified as unknown.
This could be because these classes are not tightly clustered in the feature space of the classifier, causing the signals at the edges of the clusters to be classified with lower confidence. 
The unknown AM and FM signals have the highest rejection accuracy, at 100\%. This could be because they are the only signals that use analog modulation, making them the most separated from the digital modulations used in the known classes. The generic communication waveforms, OFDM, SC-FDMA, and SC, have slightly lower accuracies. This could be because they are more similar to the waveforms used in known classes.

\section{Conclusion}

In this paper, the efficacy of an open set wireless standard classifier was demonstrated. When SNRs were greater than 0 dB, the open set classifier achieved average accuracies of 86\% and 95\% for the known and unknown classes respectively. These results are in spite of signal impairements, including multipath fading, AWGN, frequency offsets, and I/Q imbalances.

The open set case presents a significantly more difficult scenario, which is why the accuracy is lower than the closed set case. However, the open set classifier is a much more practical system for real-world situations, where all signals cannot be guaranteed to be from known classes. The unknown detection algorithm proposed adds a minimal amount of complexity, making it computationally practical to add to deep learning classifiers.


An area of open set classification that requires more investigation is the impact of the feature space learned by the classifier on the unknown class detection ability. If the classifier does not learn a feature space that generalizes well to unknown signals, the model will not be able to differentiate unknown signals from known signals. In general, increasing the number of known classes the model is trained with should improve the feature space. However, determining when a feature space is good enough is key to guaranteeing performance with any unknown classes.

\bibliographystyle{ieeetr}
\bibliography{ref.bib}

\end{document}